\documentclass[aps,showpacs,pra,twocolumn]{revtex4}
\usepackage{epsfig,amssymb,amscd}

 \newcommand{\be}{\begin{equation}}
 \newcommand{\ee}{\end{equation}}
 \newcommand{\bea}{\begin{eqnarray}}
 \newcommand{\eea}{\end{eqnarray}}

 \newcommand{\C}{{\mathcal C}}

 \newcommand{\A}[1]{{\mathcal A}_{#1}}

\def\Nc60{${}^{15}$N@C60}
\def\Pc60{${}^{31}$P@C60}

\def\A{{\cal A}}
\def\B{{\cal B}}

\def\C{{\cal C}}
\def\D{{\cal D}}

\def\X{{\cal X}}

\def\Iza{\hat{I}_z^{\cal \,A}}
\def\Sza{\hat{S}_z^{\cal\,A}}
\def\Izb{\hat{I}_z^{\cal\,B}}
\def\Szb{\hat{S}_z^{\cal\,B}}
\def\SWAP{SW\!\!AP}

\begin{document}


\title{ Quantum cellular automata quantum computing with endohedral fullerenes}
\author { J. Twamley\footnote[2]{Email: jtwamley@thphys.may.ie} }
                
\affiliation{Department of Mathematical Physics,\\
National University of Ireland Maynooth,\\ Maynooth, Co. Kildare, Ireland}
\received{\today}

\begin{abstract}
We present a scheme to perform universal quantum computation using global addressing techniques 
as applied to a physical system of  endohedrally doped fullerenes. 
The system consists of an $\cal ABAB$ linear array of Group V endohedrally doped fullerenes.
Each molecule spin site consists of a nuclear spin coupled via a Hyperfine 
interaction to an electron spin. The electron spin of each molecule is in a quartet ground state 
$S=3/2$.
Neighboring molecular electron spins are coupled via a magnetic dipole interaction. 
We find that an all-electron 
construction of a quantum cellular automata is frustrated due to the degeneracy of the 
electronic transitions. However, we can construct a quantum cellular automata quantum 
computing architecture using these molecules by encoding the quantum information on the 
nuclear spins while using the electron
spins as a local bus. We deduce the NMR and ESR pulses required to execute the basic cellular 
automata operation and obtain a rough figure of merit for the the number of gate operations
per decoherence time. We find that this figure of merit compares well 
with other physical quantum computer proposals. We argue that the proposed architecture meets well
the first four DiVincenzo criteria and we outline various routes towards meeting the 
fifth criteria: qubit readout.
 \end{abstract}
\pacs{03.67.-a, 03.67.Lx, 73.21.-b, 73.22.-f, 76.70.Dx}
\maketitle

\section{Introduction}
Quantum information processing (QIP), relies on the ability to perform a selection of 
unitary operations on a multi-partite system. More precise criteria that must be met by 
any physical implementation of a quantum information processor have been proposed 
in \cite{DiVen1}. In particular, in a spin-based quantum computer architecture the 
spin sites (i) should be easy to  physically manipulate, 
(ii) easy to replicate in numbers, (iii) they should possess some type of inter-site 
interaction and (iv) they also should be somewhat isolated from their environment. 
Meeting such ideals in practice is very demanding. 
It is believed by some that a solid-state implementation could prove highly advantageous in terms 
of the scaling up of the number of spin sites and many  solid-state proposals for a quantum 
information processor are currently under study \cite{Kane98,Spiller01,review}. However, to
achieve (iv), usually requires the spin-sites to be located within a near-perfect crystal
lattice and this leads to great difficulties in satisfying (i) and (ii) above.
If the interaction between spin sites is mediated by a predominantly electronic interaction, as 
is the case in ion-trap quantum computers and \cite{Kane98}, the quantum
computer operation will be very sensitive to stray electric charges. Such systematic 
charge noise can be source of heating and may be very difficult to eliminate.
Additionally the use of local qubit addressing is ubiquitous in very many quantum computer designs. However
the use of such local qubit addressing comes with a potentially very large overhead such as the 
building of nanoscopic metal gates as in \cite{Kane98}, or the greatly 
increased bandwidth required in NMR quantum 
computing in order to frequency address a large number of spins in a molecule \cite{Jones01}. 
For small numbers of qubits,
this overhead may be acceptable but the decoherence effects of imperfect local addressing in a device 
containing many tens of qubits may be much more serious. 
Finally, the use of nuclear spins alone in a spin-based quantum computer design leads to
a serious difficulty in achieving a sufficiently high initial spin polarization with which to
initialise the quantum computer. This is a primary obstacle in liquid state NMR quantum computing
and the problem remains in solid-state designs even when the nuclei are cooled to milli-Kelvin temperatures.

Faced with all these challenges, a number of us \cite{QIPDDF,Harneit00,LimandSuter}, following ideas 
put forward by the author in \cite{QIPDDF},  have
considered a molecular-based quantum computer which may offer 
a potentially more robust 
packaging of the qubit.  The central idea is to encode a qubit into the spin system
contained {\em within} 
 an endohedrally doped fullerene. 

In earlier publications \cite{Harneit00}, 
we outlined the unusual properties of Group V C60 endohedral fullerenes 
and indicated their potential 
use in a solid-state implementation. We, however, did not there give a detailed description 
of a concrete quantum computing architecture for use with this material.

In \cite{LimandSuter}, a locally-addressed quantum computer design using endohedral fullerenes
was proposed. Briefly this design stored the qubit on the nuclear spin of the endohedral atom 
and used the endohedral electron
spins as a local bus. The interaction between neighboring qubits is 
mediated by the magnetic dipole force between 
neighboring endohedral electrons. 
By constructing a linear chain of molecules and subjecting the chain to a large
magnetic field gradient one can frequency address the electronic spin of an 
individual molecule and through this perform universal quantum computation. 

This type of design has a number of serious drawbacks: (a) to achieve fast local addressing the 
resonant frequencies of neighboring endohedral electrons must be well separated. Thus the 
microwave bandwidth required to frequency address many frequency separated spins 
grows with the number of qubits and may become technically quite challenging to achieve  
beyond a few hundred MHz (which corresponds to a few tens of qubits). 
In addition, (b) to be scalable  one must perform quantum error correction and 
to do so will require the simultaneous performance of various
quantum gates within the device. Thus the number of independent 
spectrometer frequency channels required to produce such multi-frequency microwave pulses will
grow with the device. To engineer such a spectrometer may be not be ultimately scalable.
Finally, (c) to achieve local addressing in \cite{LimandSuter} requires the generation of 
a magnetic field which is highly 
stable in time and possesses a very large, and preferably very homogeneous, spatial gradient. 
Currently spectrometers are engineered to yield highly homogeneous, ultra-stable
constant magnetic fields (zero spatial gradients), 
and the engineering of magnetic fields desired in \cite{LimandSuter}, 
on a microscopic scale, though possible, will require further development.
 
The purpose of this work is to show that a quantum cellular automata 
quantum computing architecture (QCAQCA), can successfully be applied to operate 
with Group V C60 endohedrals. Further, the resulting design does not exhibit
the above mentioned drawbacks seen in \cite{LimandSuter}. The QCAQCA design 
only requires an homogeneous constant magnetic field. More importantly, 
in the QCAQCA, full scale universal quantum computation with quantum error correction only requires
frequency addressing at a fixed, small number of frequencies, independent of the size of 
the device. These two advantages mean that a quantum cellular automata based device
may be far easier to develop in hardware than any locally addressed architecture. 

For the design developed below we exhibit the specific pulse primitives necessary for
quantum cellular automata quantum computation. We are also able to estimate 
rough figures of merit, or number of quantum gates that can be executed in a decoherence time.
The estimated figure of merit obtained below compares well with other proposed 
quantum computer designs. 

\vspace{.5cm}
In section II below we review the properties of ${\cal X}@$C60 (${\cal X}={}^{15}N,\;{}^{31}P)$, 
pertinent for quantum information processing. In section III we briefly outline the 
essential ingredients of QCAQCA. In section IV we develop a number of tools (pulse sequences), 
for manipulating the Hamiltonian of an alternating linear chain, $\cal ABABAB$, of these endohedrals 
where ${\cal A}=$\Nc60 and ${\cal B}=$\Pc60. 
In section V we examine the possibility of an all-electronic QCAQCA and find that universal quantum computation seems not possible using the electrons 
alone. In section VI we consider an architecture which also includes the nuclear spins and 
show, by developing new gate pulse sequences, that universal quantum computation can be 
achieved via a quantum cellular automata design. We finally, in section VII, propose a 
number of readout techniques that might be applicable to the proposed ensemble 
(a Type-II quantum computer design \cite{Yepez}), and single-issue 
(or a Type-I quantum computer design \cite{Yepez}), quantum computer design.

 \section{Group V Endohedrals}
\label{endos}
The endohedrals \Nc60, \Pc60 and N@C70 all exhibit very 
sharp ESR spectra \cite{Weidinger98,Knapp98}. This indicates the presence of free 
electrons within these molecules. It has been further shown, both theoretically and 
experimentally \cite{Greer00, Dinse00, Smith01}, that the trapped atom ($N$ or $P$), 
sits at the geometric center of the fullerene cage and that the electrons are in a $S=3/2$ 
quartet ground state. The trapped atoms are extreme examples of compressed atoms 
\cite{Buchachenko01, Greer00}, where the electronic wavefunction of the trapped atom 
is repulsed away from the encompassing carbon cage and suffers a spatial compression. 
This unique type of ``bonding'' leads to significantly higher nuclear-electron 
wavefunction overlap and thus larger Hyperfine coupling constants than what is 
found in ``free'' atoms. For the case of  C60 Group V endohedrals, the large 
symmetry renders this hyperfine coupling highly isotropic. 
Another consequence of this compression is the virtual lack of any electronic interaction 
between the trapped atom and the carbon cage. The trapped atom is motionally 
confined by a harmonic-like potential to the center of the cage. 
In almost all respects, the trapped atom behaves as a ``free'' (unbonded) atom, though 
spatially restricted to be within the fullerene cage. All of the above findings 
imply that such endohedrals behave as {\em nanoscopic molecular neutral atom traps}. 

The distributed $\pi$ bonding electrons on the C60 also act as an almost perfect 
Faraday cage, strongly isolating the electrons of the trapped atom from external 
electric fields \cite{Pietzak97}. This latter observation implies that it would 
require the application intense local electric field gradients $\sim 1V/nm$, in 
order to alter the Hyperfine coupling constant of the trapped atom. Such an 
electrostatic addressing scheme has been proposed to address nuclear spin qubits in the 
Phosphorus donors of the Kane design \cite{Kane98}, but with far lower field gradients. 
Such a scheme would prove very difficult to execute here due to the requirement 
of such intense electric field gradients. One very significant advantage of 
containing the spin site inside a relatively large C60 molecule (diameter $\sim 1nm$), 
is that such endohedrals can be nano-positioned using current scanning tunneling 
microscope techniques \cite{Beton00}, and moreover neighboring C60 molecules can be 
nano-positioned on the Silicon(100)-2x surface to be as close as $\sim1.1nm$. 
The C60 molecules, on the Silicon(100)-2x  surface, are fixed translationally at 
room temperature and also rotationally at lower temperatures. More speculatively, 
due to the electronic-wavefunction spatial compression experienced by these molecules, 
the chemistry of the doped Group V fullerene material is practically identical to 
undoped fullerene. It may then be possible to self-assemble large organised spin 
structures using existing well-studied synthetic supra-chemistry techniques \cite{Harneit00}.  

The very sharp ESR spectra from these molecules indicates very long relaxation times and 
more recent measurements have shown the electronic relaxation times are $T_1\sim 1s$ at 
$T\sim 7^\circ K$, while $T_2\sim 20\mu s,\;\;\forall \,T$ \cite{Knor00}. No nuclear 
relaxation times have yet been recorded but they are expected to be several orders of 
magnitude longer than the electronic relaxation times. The measured $T_2$ time contains 
contributions from several sources, e.g. unresolved dipolar couplings. The theoretical maximum of $T_2$, in the complete absence of
the unwanted interactions with other paramagnetic impurities, 
is the relaxation time $T_1$. Indeed, for the case of Phosphorus defects in isotopically
 ultra-pure 
${}^{28}$Si, the phase relaxation time of the loosely bound Phosphorus electrons can be 
as long as $T_2>.1$ msec with $T_1>1$ hour \cite{Phase}. There $T_2$ 
is limited by Hyperfine interactions with residual
${}^{29}$Si nuceli \cite{Kane00}. The indicated 
increase of $T_2$ towards $T_1$ in the situation of vanishing spin density
is crucial for all current spin-based quantum computer proposals 
\cite{Kane98,review,Spiller01,QIPDDF,Harneit00,LimandSuter,Phase1}. Achieving such limits will be very challenging. Making these assumptions
for the endohedral electrons we expect $T_2\sim T_1\sim 1$sec, at $T\sim 7^\circ K$,  
with perhaps longer times at lower temperatures.
Such Electronic $T_1$ relaxation times ($\sim 1$sec),
 in such complicated molecules is highly unusual and may be unique in all
non-crystal hosted spin-sites. Long relaxation times also add to the usefulness 
of such material as hosts for storing and manipulating quantum information. 

An essential ingredient for any quantum information processing is an inter-qubit 
interaction which can generate entanglement. For Group V C60 endohedrals this 
cannot be a direct electronic exchange type interaction as the electronic 
wavefunctions are tightly compressed to be within the C60 cage. The C60's 
Faraday cage does not restrict magnetic interactions and neighboring endohedrals 
experience a significant magnetic dipole coupling, 
$H_D\sim\tilde{J}(1-3\cos^2\theta)[3\hat{I}_z\hat{S}_z-\hat{\bf  I}\cdot\hat{\bf S}]\sim J_D\,\hat{I}_z\hat{S}_z$, 
in the weak coupling limit. The strength of this dipolar coupling has been 
estimated to be $J_D\sim 50$MHz$ (1/r^3)$, where $r$ is the separation between 
the neighboring trapped atoms in the endohedrals measured in nanometers \cite{Waiblinger00}, 
and $\theta$ is the angle  between $\vec{r}$ and the external $\vec{B}$ field. 
The Hyperfine coupling constants for \Nc60, \Pc60 have been measured \cite{Waiblinger}, 
and are given in Table \ref{table1}. 

\begin{figure}[p]
\begin{center}
\setlength{\unitlength}{1cm}
\begin{picture}(6,13)
\put(-1,-0.5){\includegraphics[width=7cm]{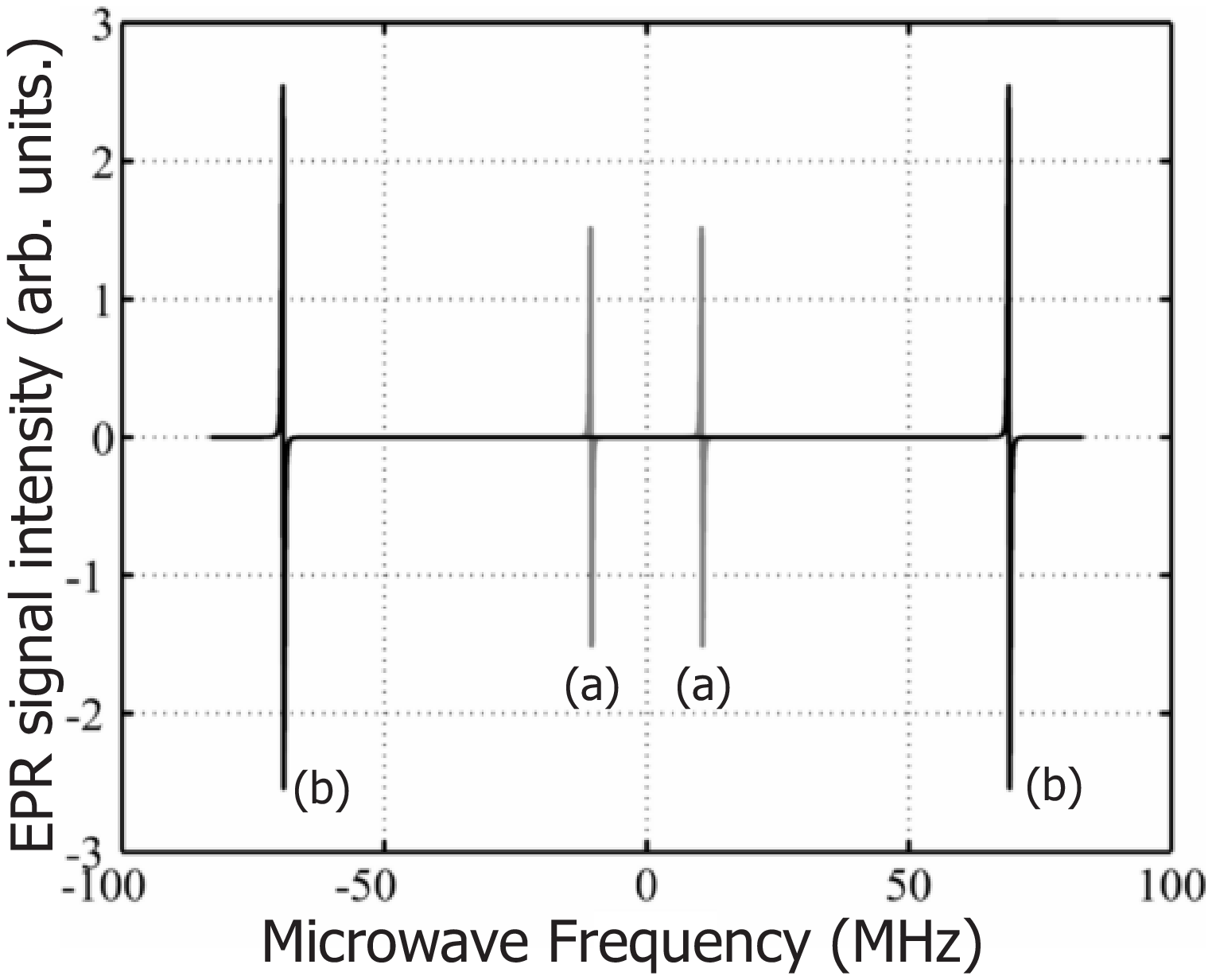}}
\put(-1.3,6){\includegraphics[width=7.25cm]{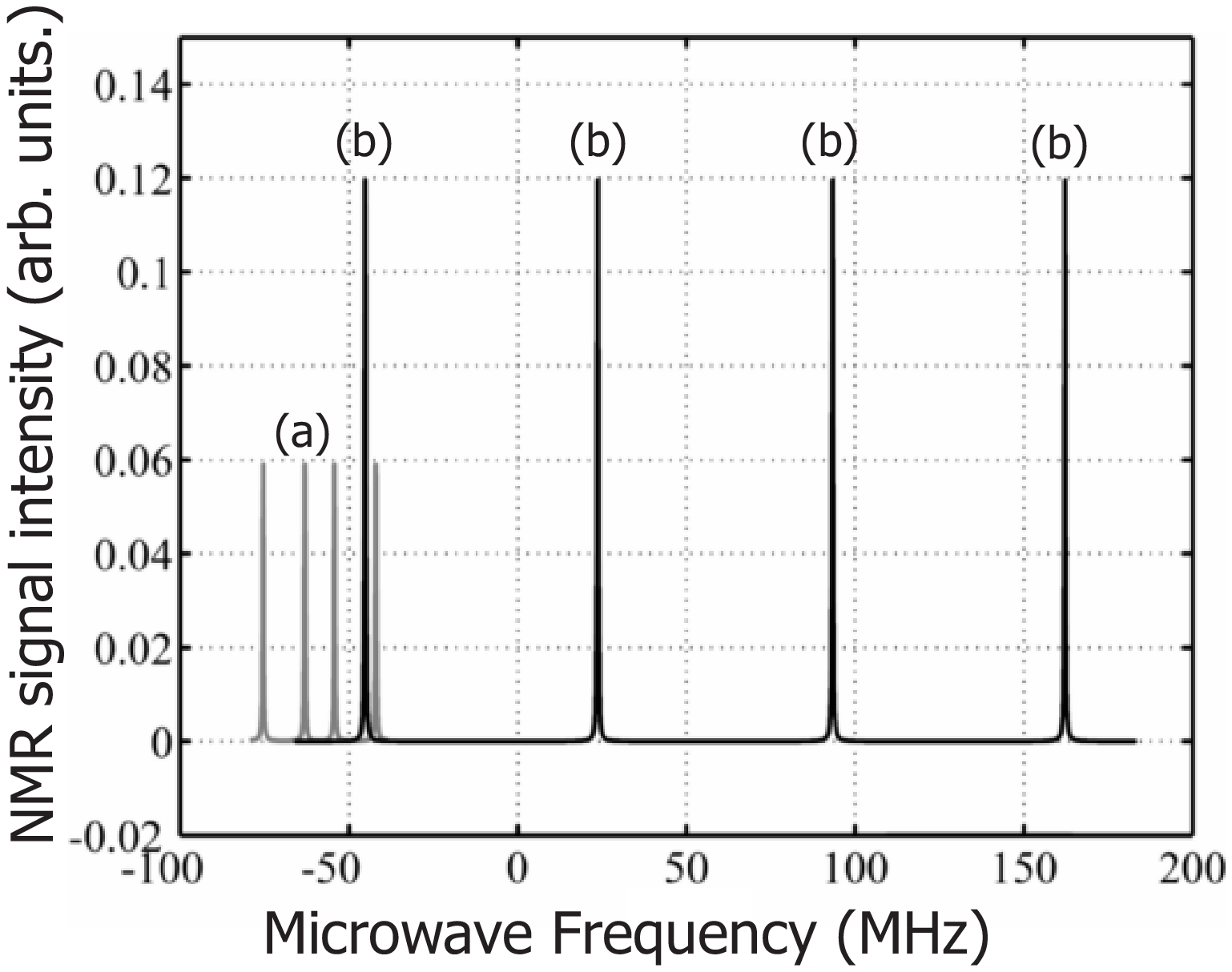}}
\put(-1,5.5){{\Large (B)}}
\put(-1,12){\Large (A)}

\end{picture}
\end{center}
\caption{Simulated NMR (in subfigure A), and ESR (in subfigure B), spectra of $\A=$\Nc60 and $\B=$\Pc60 with (a) and (b) labeling the spectral lines from $\A$ and $\B$ respectively. }
\label{fig1}
\end{figure}

From the above, to a first approximation, the full 
spin Hamiltonian for the pair $\cal AB$, (${\cal A}=$\Nc60, ${\cal B}=$\Pc60), can be given as
\begin{eqnarray}
\hat{H}/h&=& g_e\mu_eB_z \Sza-g_N^{\cal A}\mu_NB_z\Iza+A^{\cal A}\Sza\Iza\nonumber\\
&+&g_e\mu_eB_z\Szb-g_N^{\cal B}\mu_NB_z\Izb+A^{\cal B}\Szb\Izb\nonumber\\
&+&J_D\Sza\Szb\
\label{Ham1}
\end{eqnarray}
where $I(S)$, labels nuclear(electronic) spin, and we have made the secular 
approximation dropping terms which do not commute with the electronic Zeeman 
Hamiltonian since $\omega_{Larmour}^S\equiv g_e\mu_eB_z \gg (J_D,\,\omega_{Larmour}^I\equiv 
g_N\mu_NB_z ,\,A)$. 
The individual ESR and NMR spectra of these molecules are simulated in Figure \ref{fig1}, 
and have been measured in \cite{Knapp98}.

\section{Quantum Cellular Automata}
Typical quantum computer architectures assume the capability of locally 
addressing every qubit. Such a requirement is highly challenging to engineer 
and the achievement of this level of individual quantum control is a major goal 
in almost every current implementation. The use of such local control methods forces 
an interaction of the quantum information with very many ``classical'' control gates, 
each possibly providing a decohering effect on the quantum information. It is also 
important that the effects of such control gates on the computer's Hamiltonian remain 
as ``classical''  as possible in that they effect a change in the parameters appearing 
in the system Hamiltonian and do not themselves become entangled with the quantum computer. 
This criteria of classical gating becomes non-trivial as the length scales of the 
quantum computer architecture reduces to nanometers. 

An alternative architecture 
is to utilize a small number of identifiable spins, placed in a regular spatial pattern, 
and to manipulate the quantum information encoded in this spin chain via global addressing techniques. This first of such globally addressed architectures 
was invented by Lloyd in 1993 \cite{Lloyd93}, and utilized three types of addressable
spin arranged in the linear pattern $\cal ABCABCABC$, and where each spin site encodes one 
logical qubit. In fact this cellular automata design was one of the very earliest 
quantum computer architectures proposed in the literature. Lloyd showed that such a 
quantum computer architecture was universal. Benjamin \cite{Benjamin99}, found a similar 
architecture which used only two types of identifiable spin species arranged in the 
alternating linear pattern $\cal ABABAB$. This reduction in spin resources came with an 
increase in logical encoding: a logical qubit is now encoded into four spin sites with 
a buffer space of at least four empty spin sites between each logical qubit. 

The operation of Benjamin's architecture centers on the ability to implement 
the global unitary operator, $\hat{\cal A}_f^U$. Denoting the spin up(down) state as $1(0)$,
$\hat{\cal A}_f^U$ is the conditional application of the unitary $U$ to the $\cal A$ spins
in the alternating spin chain $\cal ABABAB$, depending on the state of $\cal A$'s 
neighboring $\cal B$
spins. In particular, letting $f$ be the sum of the states of the neighboring $\cal B$ spins, 
we have $f\in [0, 1, 2]$.  Thus, for example, 
$\hat{\cal A}_1^U$ is  the conditioned application of $U$ to all 
$\cal A$ spins in the alternating chain which have neighboring $\cal B$ 
spins that are different from each other, i.e. $f=1$.
One has a similar definition for 
for $\hat{\cal B}_f^U$, for the application of $U$ on all $\cal B$ spins conditioned 
on their neighboring $\cal A$ spins. The case when $U\sim{\rm NOT}$, in $\hat{\cal A}_f^U$ 
(or $\hat{\cal B}_f^U$), occurs quite frequently in Benjamin's architecture and is shortened to
$\hat{\cal A}_f$ (or $\hat{\cal B}_f$).

Benjamin shows in \cite{Benjamin99}, that through a 
judicious sequence of applications of $\hat{\cal A}_f^U$ 
and $\hat{\cal B}_f^U$,  one can implement single qubit operations and two-qubit 
CNOT operations and thus  the architecture is universal. 
In particular, to move qubits through the spin chain one applies an alternating pulse
sequence of $\hat{\cal A}_1$ followed by $\hat{\cal B}_1$,
while the generation of a control-U between two neighboring logical qubits requires $\sim 30$ 
global pulses \cite{Hovland00}. One can translate all the standard circuit-based 
quantum algorithms to run with this quantum cellular automata architecture \cite{Hovland00}. 
This programming architecture, though somewhat expensive in terms of spatial 
and temporal resources, is ideal for use in systems where the individual control 
of qubits is difficult and it could provide a interim test-bed for various 
implementations while the more long term goal of local gating is developed. 

\begin{figure}[p]
\begin{center}
\setlength{\unitlength}{1cm}
\begin{picture}(6,2)
\put(-1.7,0){\includegraphics[width=9cm]{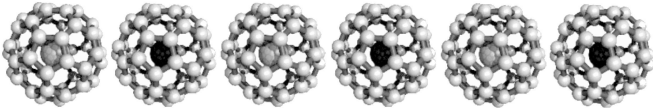}}
\end{picture}
\end{center}
\caption{Schematic depiction of the $\A\B\A\B\A\B$ quantum cellular automata chain using the Group-V endohedral fullerenes.}\label{chain}
\end{figure}

Since Lloyd's and Benjamin's pioneering work little has appeared in the literature with 
regard to quantum cellular automata designs for quantum computers \cite{Benjamin_recent}. 
In this paper
 we will consider  a quantum cellular automata 
architecture consisting of an alternating linear array of Group V endohedrals, $\cal ABABAB$, 
with the $\cal AB$ Hamiltonian given by (\ref{Ham1}) (schematically depicted in 
Figure \ref{chain}). 
Essentially this task reduces to developing a ``quantum algorithm'' 
for generating Benjamin's global operation $\hat{\cal A}^U_f$ (similarly for $\hat{\cal B}^U_f$). 
We first investigate the possibility of implementing $\hat{\cal A}^U_f$, 
using the $S$ (electron), spins alone.

\section{The global operation}
\label{global}
In this section we consider an alternating spin 1/2 chain of electronic spins 
$S^{\cal A}S^{\cal B} S^{\cal A} S^{\cal B}$. 
We begin by considering the $\cal S^{\cal A}S^{\cal B}$ unit alone where 
$\cal A$ spin has only one neighboring
$\cal B$ spin. We assume that we know how to  
 perform the controlled operation of applying $U$ to $S^{\cal A}$, with $S^{\cal B}$ 
 controlling, an operation we denote by $C({\cal B},{\cal A};U)$. 
We shall  describe below the details regarding the sequence of pulses required to execute 
$C({\cal B},{\cal A};U)$.
We next consider the case when $\cal A$ has two neighbors, e.g. 
$S^{\cal B} S^{\cal A} S^{\cal B}$. Remembering that one cannot separately address an individual
neighboring $\cal B$ spin one can show that instead of performing $C({\cal B},{\cal A};U)$, the
above mentioned sequence of pulses now performs 
\begin{equation}
\Xi({\cal B},{\cal A};U)\equiv C({\cal B},{\cal A};U)C({\cal B},{\cal A};U)=C({\cal B},{\cal A};U)^2\;\;,
\label{mychi}
\end{equation}
where Eq. (\ref{mychi}), defines the function $\Xi$. One can observe that the case when 
the desired global
operation is $\hat{\cal A}_1\equiv \hat{\cal A}_1^{\rm NOT}$, we have $\hat{\cal A}_1=\Xi({\cal B},{\cal A};{\rm NOT})$.

We now indicate the 
procedure for implementing $\hat{\cal A}^U_2$, when the desired $U\sim \exp(i\theta S_x^{\cal A})$.
For ease of notation we define the standard NMR/ESR symbol for $\pi/2$ pulses
\cite{ESRPauli,Linden99, Schulte00},
\begin{equation} 
[+S^{\cal B}_y]\equiv \exp(i \frac{\pi}{2} S^{\cal B}_y)\;\;,
\end{equation}
and the more general rotation through the angle $\gamma$ about the $z-$axis, for example, by the 
symbol
\begin{equation}
\{Z^{\cal A}_\gamma\}\equiv  \exp(i\gamma S^{\cal A}_z)\;\;.
\end{equation}
We find that we can execute $\hat{\cal A}^U_2$, when $U\sim \exp(i\theta S_x^{\cal A})$
 through the following  pulse sequence (read right to left),
\begin{eqnarray}
 \hat{\cal A}^U_2&=&\Xi({\cal B},A;\{X^{\cal A}_\delta\})\cdot \{X^{\cal A}_{\pi/2-\delta}\}
\cdot\{Z^{\cal A}_{-\pi/2}\}\cdot 
\Xi({\cal B},{\cal A};\{X^{\cal A}_\pi\}) \nonumber\\
&&\cdot \{Z^{\cal A}_{-\pi/2}\}\cdot \{X^{\cal A}_{\pi/2-\delta}\} 
\cdot \Xi(B,A;\{X^{\cal A}_\delta\})\;\;.\label{flips}
\end{eqnarray}
where $2\delta=\theta$.
To implement $\hat{\cal A}^U_1$ 
one sets $\delta=\theta$ and replaces the left most term in (\ref{flips}), 
$\Xi({\cal B},{\cal A};\{X^{\cal A}_\delta\})$, by $\{Z^{\cal A}_{\pi}\}$, 
while to implement $\hat{\cal A}_0^U$, 
one applies $\{X^{\cal B}_{\pi}\}$ to all the $\cal B$ spins, then $\hat{\cal A}_2^U$, 
and finally flip the $\cal B$ spins back with $\{X^{\cal B}_{-\pi}\}$. The above pulse sequence 
(\ref{flips}), can be easily modified to suit other forms of $U$ besides $\exp(i\theta S_x^{\cal A})$, 
however we will find below that the most useful case is when $\theta=\pi$ and
$U=\exp(i\pi S_x^{\cal A})\sim NOT$ on the $\cal A$ spin.

\section{All Electronic Quantum Cellular Automata Quantum Computer}
\label{electronic}
From Figure \ref{fig1}, we can separately address the electronic spins $S^{\cal A}$ and $S^{\cal B}$ 
with a multifrequency selective soft pulse of length $>20ns$. If we first consider
 both $S^{\cal A}$ and $S^{\cal B}$ to both be spin-$1/2$, then one can use standard pulse encodings
 of the CNOT \cite{Linden99, Schulte00},
\begin{eqnarray}
C(\A,\B,NOT)&=&[-S^\B_y][\mp S^\A_z\mp S^\B_z][\pm 2S^\A_zS^\B_z][+S^\B_y]e^{\pm i\pi/4}\;\;,\\
C(\B,\A;NOT)&=&[-S^\A_y][\mp S^\A_z\mp S^\B_z][\pm 2S^\A_zS^\B_z][+S^\A_y]e^{\pm i\pi/4}\;\;,\label{spinhalfcnot}
\end{eqnarray}
 to generate  the operation $\hat{\A}_1$ and $\hat{\B}_1$ of the previous section. 
To perform universal quantum computation one must be at least capable of implementing
a two-qubit CNOT and a minimal set of one-qubit operations such as the 
phase gate, 
the $\pi/8$ gate, and the Hadamard gate.
In the quantum cellular automata architecture of Benjamin this means that one must at least 
be able to implement the global operation $\hat{\A}_f^U$, 
where $U$ is taken from this minimal set of one qubit gates and the case where $U\sim NOT$. 
NMR and ESR pulses effectively generate terms in the Hamiltonian which are linear in the spin operators, $I_{x, y, z}$, and $S_{x, y, z}$.  In particular, the spin operators when commuted close  to form the Pauli algebra. In the case of spin-1/2, the Pauli algebra generates the group SU(2), or all possible spin-1/2 unitaries.  Thus one can generate any desired spin-1/2 unitary through a sequence of pulses when acting on an isolated spin. Furthermore, in the case of the alternating chain $\A\B\A\B\A\B$, one 
can find suitable alterations of (\ref{spinhalfcnot}), when used in (\ref{mychi}) and (\ref{flips}), to generate any given unitary $U$ in $\hat{\A}_f^U$ and similarly for $\hat{\B}_f^U$.
Thus an all electronic quantum cellular automata quantum computer architecture  (QCAQCA),
 using a spin-$1/2$ $\A\B\A\B\A\B$ chain is possible. 

The  
endohedrals \Nc60 and \Pc60, however posses electron spin $S=3/2$ and the application of Benjamin's QCAQCA is not as straightforward. One possibility is to consider the four electronic levels of the electron spin with $S=3/2$ to constitute a qu-dit  with $d=4$. Multi-level quantum computation has been proposed elsewhere \cite{multiqdit}, but to extend Benjamin's formulation of the spin-1/2 QCAQCA to operate with qu-dits is far beyond the scope of this article. Indeed, as shown below there are reasons to believe that that such an extension may not be possible within the confines of NMR and ESR. 

Remaining with qubits 
we instead propose to encode 
a qubit into the four levels of a spin $3/2$ in two ways: an ``inner qubit'' 
$\{ |m_s=\pm 1/2 \rangle\}$, and an ``outer qubit'' $\{|m_s=\pm 3/2\rangle\}$. 
By a small alteration of (\ref{spinhalfcnot}), one can obtain
\begin{widetext}
\begin{eqnarray}
C_{inner}(S^\A,S^\B;NOT)=C_{outer}(\overline{S_\A},S^\B;NOT)&=&[S^\B_y][- S^\A_z- S^\B_z][+2S^\A_zS^\B_z][-S^\B_y]e^{+ i\pi/4}\;\;,\label{spinthreehalfcnotinner}\\
C_{inner}(S^\B,S^\A;NOT)=C_{outer}(\overline{S_\B},S^\A;NOT)&=&[S^\A_y][- S^\A_z - S^\B_z][+ 2S^\A_zS^\B_z][-S^\A_y]e^{\pm i\pi/4}\;\;,\label{spinthreehalfscnot}
\end{eqnarray}
\end{widetext}
where $C_{inner}$ is a $CNOT$ on the Hilbert space spanned by the inner qubits, 
and similarly for $C_{outer}$, $\overline{S^A}=NOT(S^A)$, and we have used the spin-$3/2$ 
representations of the $SU(2)$ group. Armed with this one can easily construct the 
Benjamin global operations $\hat{A}_1$ and $\hat{B}_1$ on the inner qubit subspace. 
One can 
further construct (see section \ref{tagging}), the global operations $\hat{A}_f^{NOT}$, 
(and similarly $\hat{B}_f^{NOT}$), for $f=0,\,1$ and $f=2$, the operations most frequently
 used in the quantum cellular automata quantum computation. 

However one finds that one cannot generate all the one-qubit gates required for universal quantum computation. As above,  ESR pulses essentially generate terms in the Hamiltonian that are linear in the spin operators and again these terms commute to form a Pauli algebra. Evolution under such terms lead to dynamics that resides in the group generated by this Pauli algebra (or $SU(2)$). However, since now we have $S=3/2$, this $SU(2)$ is merely a subgroup of the full group of unitaries that can act on a spin-3/2, but as long as the system Hamiltonian is linear in the spin operators we need only to consider the dynamics within the $SU(2)$ subgroup. 
Normally a unitary operation $U\in SU(2) \subset SU(4)$ will not factor into unitary operations on the inner and outer qubit subspaces. However if $U$ is diagonal in the $|m_s\rangle$ basis it can so factor and with this one can find pulse sequences to implement $\hat{A}^U_f$ where $U$ is a phase gate, and the $\pi/8$ gate but crucially not the Hadamard gate. Thus an all electronic-QCAQCA with a system Hamiltonian which is linear in the spin operators and where $S>1/2$ seems not possible.

One way of implementing
$SU(2)$ operations on the qubit subspaces and one which is well known in NMR and ESR is to introduce terms in the system Hamiltonian which are not linear in the spin operators. Such terms lie in the full spin-3/2 algebra and the generated unitaries no longer lie in the $SU(2)$ subgroup but in the much larger $SU(4)$ group. One effect of such nonlinear terms, for example Zero-Field-Splitting terms: $H_{ZFS}\propto S_z^2$, can be to lift the degeneracy of the microwave transition frequencies between the  $|m_s\rangle \leftrightarrow |m_s+1\rangle$. This allows one to frequency address the transition $|m_s=-1/2\rangle \leftrightarrow |m_s=+1/2\rangle$, or the inner qubit subspace, and directly implement $SU(2)$ operations on this subspace. This allows the generation of arbitrary one-qubit unitaries $U$ \cite{Wokaun77}. However, this solution frustrates the coherence transfer portion, $[+2S^\A_zS^\B_z]$, of the pulse sequence (\ref{spinthreehalfcnotinner}), and (\ref{spinthreehalfscnot}), and thus the construction of  the global QCAQCA operator $\hat{\A}^U_f$. This can be seen by realising that to obtain a $[+2S^\A_zS^\B_z]$ pulse one must effectively cancel out all the terms in the system Hamiltonian bar the term $S^\A_zS^\B_z$. As standard in NMR and ESR one does this cancellation through ``average Hamiltonian'' theory \cite{average}, via the application of various $SU(2)$ unitaries by RF or MW pulses. As long as the system evolution remains within the $SU(2)$ subspace such an averaging out can be achieved. However when terms nonlinear in the spin operators are introduced into the system Hamiltonian no  application of $SU(2)$ unitaries can, in general, 
average out the resulting ``nonlinear'' evolution (i.e. in the average Hamiltonian theory nonlinear terms in the Hamiltonian will almost always average to other nonlinear terms), and thus the generation of $[+2S^\A_zS^\B_z]$, and from this $\hat{A}^U_f$, is not possible. 

Thus we have found that when operating on  an $\A\B\A\B\A\B$, spin-$3/2$ chain with ESR pulses one cannot 
generate all of the necessary global operations  required for universal quantum 
computation.
We next find that the combination of nuclear and electronic 
spins within the chain can improve matters.

\section{Nuclear-electron quantum cellular automata}\label{NEQCA}
 \Nc60 and \Pc60  both have nuclear spin-$1/2$. In this section we assume that the 
quantum information is stored in the nuclear spins of the dopant atoms in the 
$\A\B\A\B\A$ chain. This has the significant advantage 
that the nuclear spin relaxation times are typically longer 
than the electronic spins by several orders of magnitude. As the nuclei only are coupled via the
Hyperfine interaction to the electrons, one can, via a suitable RF and MW pulse sequence, cancel out
this Hyperfine interaction with high precision. This allows the nuclei to act as a quantum
memory and store the quantum information for very long periods of time in between processing.

To execute the Benjamin global operation $\hat{\A}^U_f$ we essentially use the inner electronic qubits as a local ``bus'' for the quantum information stored in the nuclei. 
The procedure can be thought of as a quantum algorithm for
 the generation of the global operation and consists of several steps which are pictorially represented in Figure \ref{Ben_movie}.

The algorithm starts with the inner electronic qubit bus in
 the ground state ($|m_s\rangle=|-1/2\rangle$ for all molecules), 
and this bus is returned to the ground state after implementing the global
 operation $\hat{\A}_f^U$. We first 
(a) assume that  we have an arbitrary pattern of quantum information written onto the nuclear spins of the $\A$ and $\B$ molecules with the electrons set as above. 
We then, (b) swap the quantum information of all the $\B$ molecules from their nuclei to their 
inner electronic qubits. We then (c) tag those $\A$ molecules which will receive the unitary operation $U$ in $\hat{\A}^U_f$ by flipping the state of the electronic qubit of all the $\A$ molecules conditioned on the state of it's neighboring $\B$ electronic qubits.
We then (d) undo (b) by swapping the quantum information back into the $\B$ nuclei from their electrons. 

At this 
point all the inner electronic qubits are back in their ground states except for the 
``tagged'' $\A$ molecules. We then, (e) perform a 
controlled-$U$ operation on the nuclear qubits of {\em all the molecules} using their
 inner electronic qubits as the control. Finally (f), we undo the operations (d), (c) and (b).
 The system is then ready for the execution of the next global operation. 

One restriction we make in the following is that we rule out the use of selective pulses on the nuclear spins of the Nitrogen and Phosphorus. From Figure \ref{fig1}, 
to execute such selective pulses would require RF pulses of very long duration to frequency differentiate between the NMR transitions of the Nitrogen and Phosphorus. Such slow nuclear pulse sequences would lead to very long gates times. We  will therefore not allow ourselves to use such selective nuclear pulse sequences and instead make use of fast ``hard'' nuclear pulse sequences \cite{hard}. We do allow selective  addressing of the electronic transitions of the Nitrogen and Phosphorus as these pulse sequences are much shorter in duration. With this restriction however we are still able to design all the pulse sequences that are required to carry out the steps (b)-(e), above.
The resulting combined total pulse sequence is somewhat lengthy but could be substantially shortened by applying the principles of optimal control theory \cite{glasser02}.

We now expand on the above steps (a)-(e), in detail. In section (A) below we describe the initialization of the spin-chain. In section (B) we design the  $\SWAP$ pulse sequence required for step (b). In section (C) we design the electronic pulse sequence to implement the conditional operation required in step (c), and finally in section (D) we outline the execution of the controlled-$U$ needed in step (d). In section (E) we discuss the undoing of the previous steps (b)-(d), while in section (F) we provide some estimates of the pulse durations and logical QCA gate times.

\subsection{Initialization}
The QCA architecture may be implemented on an ensemble (Type-II) or single-issue (Type-I),
 quantum computer. The achievement of high nuclear polarizations has proved to be a major 
obstacle in NMR quantum computation. It remains a source of difficulty here as well.  
Polarizing the electrons is far easier, and one can achieve a difference in the ground
 to excited electronic populations of $\epsilon\sim 0.999$, at a temperature of $1^\circ K$ 
and $B\sim 10$T. Even when multiple spins are tensored to produce a pseudo-pure state, 
with such high individual polarizations one can still achieve a pseudo-pure state with a 
purity on the order of $\epsilon\sim .998$, with $1000$ spins at these temperatures and 
magnetic fields. This polarization can be transferred to the nuclei via an INEPT pulse 
sequence \cite{Levit}. To repump the electronic polarization and achieve a subset of completely spin 
(nuclear and electron), polarized molecules one could consider two chains of equal length:
 an $\A\B\A\B\cdots$ chain and a $\C\D\C\D\cdots$ chain where all $\A$, $\B$, $\C$, and $\D$ are 
globally addressable. Since at $1^\circ$K and $B=10$T, half of the spins (the electrons), 
are completely polarised, one can consider an $\A\B\A\B\cdots \A\B\A\B\X\C\D\C\D\cdots \C\D\C\D$ super-chain 
where $\X$ has a different resonant frequency yet again. One can then use a spin cooling 
quantum algorithm  \cite{Schulman99}, to efficiently spin cool {\em all} the nuclear and 
electronic spins on one side of $\X$ (ie. $\X$ acts as a spin gate). One would then switch 
off the interaction between the two sub-chains at $\X$ and use the spin polarised half chain to perform 
the quantum computation. Such a design would be suitable for a Type-II (ensemble), quantum 
computer. Alternatively, in the case of a Type-I quantum computer one can replace the $\C\D\C\D$ 
chain with an $\A\B\A\B$ chain and place the readouts on the polarised half chain. 
More practically
 any Type-I quantum computer will require a single-spin readout of the electronic spins of 
the  endohedral atoms. Such a readout may also serve to initialize the electronic 
qubits. We discuss possibilities for a single-spin readout further below.

\subsection{Nuclear-Electronic $\SWAP$}\label{neswap}
A crucial ingredient in the above mentioned quantum algorithm to generate $\hat{\A}^U_f$ is the
Hyperfine $\SWAP$. This operation performs a logical swap between a qubit stored in the nuclear spin and a qubit stored in the electron spin. Below we find two types of Hyperfine $\SWAP$, one that performs the swap using the inner electronic qubit and one that uses the outer electronic qubit. 
For the most part we will use the former type of swap but we will find later that the latter swap, which incorporates the outer electronic qubit, may be of significant use in the problem of qubit readout and we discuss this below.

To develop the pulse sequence for the Hyperfine $\SWAP$ 
it is useful to break down the $\SWAP$ operation into three $CNOT$ operations. 
Denoting $\SWAP (I:S)$, we have, $\SWAP (I:S)=CNOT(I;S)CNOT(S;I)CNOT(I;S)$, 
where $CNOT(I;S)$ is the controlled $NOT$ for $I$ controlling $S$. 
In our case $I=1/2$ while $S=3/2$. Substituting the spin-$3/2$ representations 
for $S$ into the standard spin-$1/2$ $CNOT$ pulse sequence (\ref{spinhalfcnot}), 
fails to yield a proper $CNOT$ operation. One instead must make reference to the 
two cases of a $CNOT$ with respect to the inner and outer qubits of $S$. 
With this one can find:
\begin{widetext}
\begin{eqnarray}
CNOT_{outer}(I;S)&=&[S_y][\mp S_z\mp I_z][\pm 2S_zI_z][-S_y]e^{\pm i\pi/4}\;\;,\nonumber\\
CNOT_{outer}(S;I)&=&[I_y][\pm S_z\pm I_z][\pm 2S_zI_z][-I_y]e^{\pm i\pi/4}\;\;,\label{spinhalfthreehalfscnotouter}
\end{eqnarray}
while
\begin{eqnarray}
CNOT_{inner}(I;S)&=&[S_y][\mp S_z\mp I_z][\pm 2S_zI_z][-S_y]e^{\pm i\pi/4}\;\;,\nonumber\\
CNOT_{inner}(S;I)&=&[-I_y][\mp S_z\mp I_z][\pm 2S_zI_z][I_y]e^{\pm i\pi/4}\;\;.\label{spinhalfthreehalfscnotinner}
\end{eqnarray}
With these one can derive
\begin{eqnarray}
\SWAP_{inner}(I:S)&=&[I_y][+2I_zS_z][S_y-I_y][S_x-I_x][+2I_zS_z][I_x-S_x][+2I_zS_z][I_y]e^{-i\pi/4};\;,\label{spinhalfthreehalfswap1}\\
\SWAP_{outer}(I:S)&=&[-I_y][+2I_zS_z][-I_y-S_y][-I_x-S_x][+2I_zS_z][S_x+I_x][+2I_zS_z][-I_y]e^{+ 3i\pi/4}\;\;,\label{spinhalfthreehalfswap2}
\end{eqnarray}
\end{widetext}
and the related pulse sequences are graphically shown in Figure \ref{swappulse}. 
We note for later the curious case of 
(\ref{spinhalfthreehalfswap2}), where we have in some sense ``amplified'' the 
magnetic signature of the qubit from a $\Delta m_I=\pm 1$, transition to a 
$\Delta m_S=\pm 3$, transition. In the following we will restrict ourselves to 
the inner electronic qubit subspace and thus (\ref{spinhalfthreehalfswap1}).

Having found a $\SWAP$ operation between the nuclear and inner electronic qubits of 
either molecule we now must show how one can adapt the sequence (\ref{spinhalfthreehalfswap1}),
 to swap out only the $\B$ qubits while leaving the $\A$ nuclear qubits alone. We now 
consider the case of two coupled $\A\B$ molecules with the Hamiltonian (\ref{Ham1}). 
The desired selective swap action can be achieved by replacing the  $[+2I_zS_z]$ 
terms in (\ref{spinhalfthreehalfswap1}), with the pulse sequence shown in Figure 
\ref{decouplepulse}. This new pulse sequence effectively averages out the Hyperfine coupling interaction in the $\A$ molecules thus turning off the action of the Hyperfine $CNOT$ and $\SWAP$ gates for these molecules. The Hyperfine interaction for the $\B$ molecules is left intact by the pulse sequence but is reduced in magnitude thus lengthening their Hyperfine gate durations. 

This particular pulse sequence has a number of advantages, 
the most important of which is that we do not selectively address the nuclear 
spins and thus can apply fast hard pulses to $I^\A$ and $I^\B$ simultaneously. 
We only take advantage of the selective addressing of the electronic spins of $\A$ 
and $\B$ as outlined at the beginning of section \ref{electronic}. 
Alternatively, one can use soft multi-frequency nuclear pulses in the following to 
approximate the hard nuclear pulse. In more detail, from \cite{Guenter01,Beth01}, the action
 of this pulse sequence is to remove to all orders  (within the secular approximation
 made in \ref{endos}), all interactions between $S^\A \leftrightarrow S^\B$ and $S^\A \leftrightarrow I^\A$. 
This sequence also removes all the Zeeman terms in the $\A\B$ Hamiltonian and one is thus 
left with $\hat{H}/h\sim A^{\cal B}\Szb\Izb$. When inserted into (\ref{spinhalfthreehalfswap1}), 
the effect is to swap the qubit that was stored in $I^\B$,  into the inner electronic qubit of $S^\B$. 
As mentioned above, the quantum information stored in $I^\A$ is not swapped into $S^\A$, but these spins do receive 
the local unitary transformation, $[S^\A_y][I^\A_y]$. The local unitary $[S^\A_y]$ is removed 
with a spin selective electronic pulse $[-S^\A_y]$, however we cannot remove the $[I^\A_y]$
 without performing a spin selective nuclear pulse, an operation we wish to avoid. We  
will see  however, that this extraneous local operation on $I^\A$ is not important for the 
tagging of the $\A$ molecules and furthermore the operation will be ``undone'' in step (d) (c.f. Figure \ref{Ben_movie}).

\subsection{Electronic tagging of $f$}\label{tagging}
\begin{figure}[p]
\begin{center}
\setlength{\unitlength}{1cm}
\begin{picture}(6,10)
\put(0,0){\includegraphics[width=7cm]{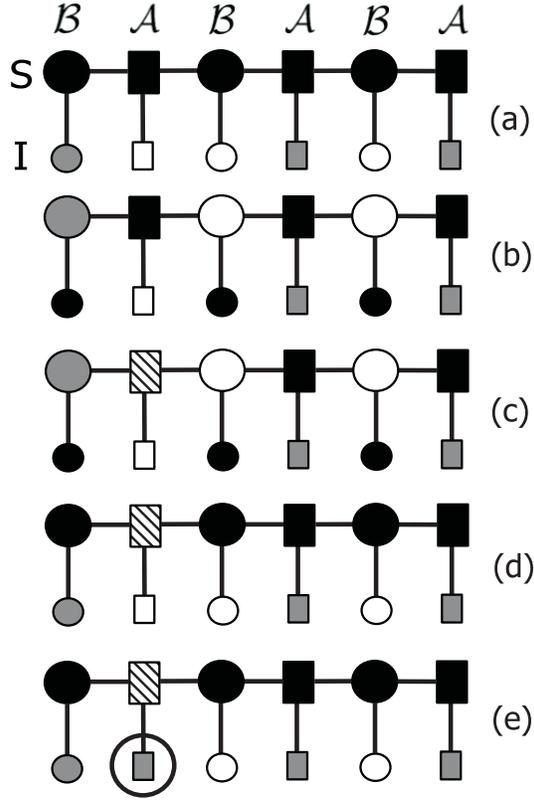}}
\end{picture}
\end{center}
\caption{Schematic representation of the steps (a)-(e) described in section \ref{NEQCA} to execute the global operation $\hat{\A}^U_f$. The alternating molecular chain $\B\A\B\A\B\A$ is depicted with the inner electronic or $S$ qubits (larger shapes), connected together via the magnetic dipole interaction, while the nuclear or $I$ qubits (smaller shapes), are connected to the electronic qubits via the Hyperfine interaction. The subfigures depict: (a) The $S$ qubits  in their ground state with some pattern of $I$ qubits; (b) A Hyperfine $\SWAP$ of the quantum information in the $\B$ molecules; (c) Conditional flipping of the $\A$ electronic qubit depending on the neighboring $\B$ electronic qubits; (d) Hyperfine $\SWAP$ of the $\B$ qubits back into the nuclei; (e) Controlled-$U$ applied to the nuclei of all molecules conditioned by the state of their electronic qubits. Here the effect of $\hat{\A}^U_f$ is to flip the state of the leftmost $\A$ nuclear qubit.}
\label{Ben_movie}
\end{figure}

\begin{figure}[p]
\begin{center}
\setlength{\unitlength}{1cm}
\begin{picture}(6,9)
\put(-3,0){\includegraphics[width=12cm]{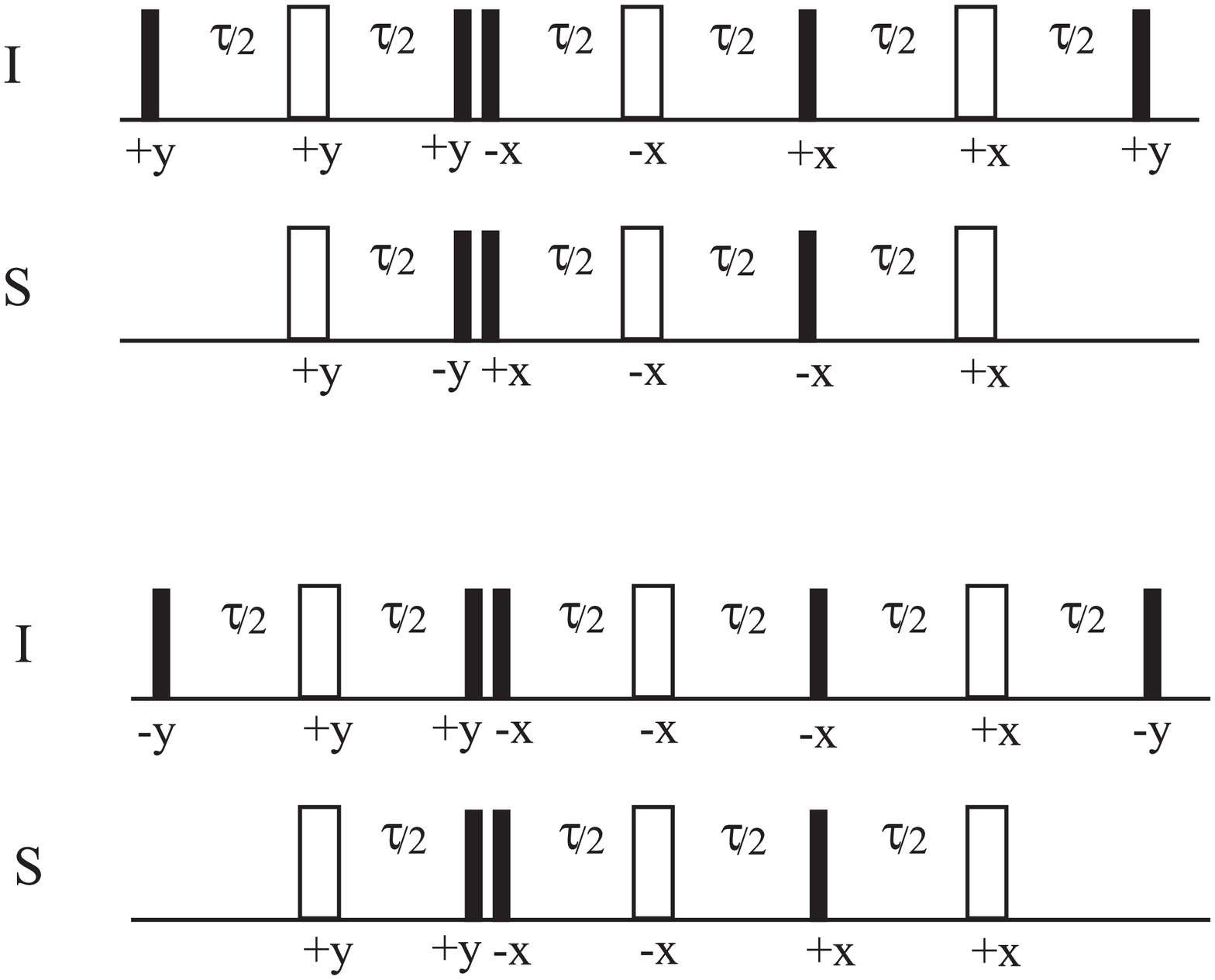}}
\put(-2.8,4){{\Large (B)}}
\put(-2.8,9.8){\Large (A)}
\end{picture}
\end{center}
\caption{Pulse sequence for a logical SWAP between spin $I=1/2$ and the inner (subfigure (A)), and outer qubits (subfigure (B)), of spin $S=3/2$ as described in (\ref{spinhalfthreehalfswap1}) and (\ref{spinhalfthreehalfswap2}). The solid black bars are $\pi/2$ pulses while the white bars are refocusing $\pi$ pulses.}\label{swappulse}
\end{figure}

\begin{figure}[p]
\begin{center}
\setlength{\unitlength}{1cm}
\begin{picture}(6,13)
\put(-3,0){\includegraphics[width=12cm]{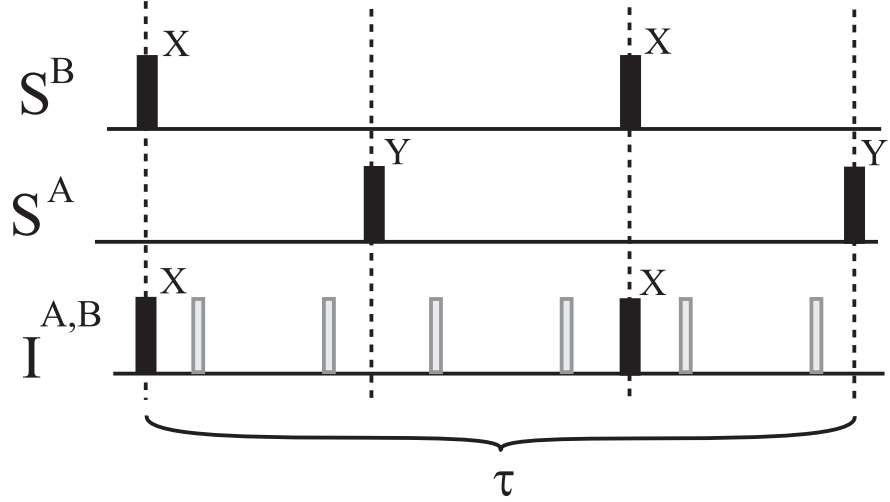}}
\end{picture}
\end{center}
\caption{Pulse sequence to selectively couple $I^\B$ with $S^\B$ while decoupling $S^\A$ with $S^\B$, and simultaneously decoupling $S^\A$ with $I^\A$. The black pulses are $\pi$ pulses and the duration of the entire sequence is matched to yield $[+2I^\B_zS^\B_z]$ in (\ref{spinhalfthreehalfswap1}). The gray pulses are fast hard $X_\pi$ pulses which are required to extend the duration of the $[2I_zS_z]$ for the case when swapping between the nuclear and electronic spin in \Pc60. By placing these pulses at the points $(m-1)/(4m)$, and $(3m+1)/(4m)$, in each third section as shown, we can extend the pulse sequence by a factor $m$. Setting $m=6$ brings the \Pc60 $SW\!AP$ to be $\sim 20$MHz, and thus long enough for the electronic spin selective pulses to operate on $S^\A$ and $S^\B$.}\label{decouplepulse}
\end{figure}

Summarizing  the previous steps: we have now swapped out the nuclear  qubits of the
 $\B$ molecules in the $\A\B\A\B\A\B\A\B$ chain into their inner electronic qubits. 
Additionally, all the inner electronic qubits of the $A$ molecules are in 
their ground states $|m_s\rangle=|-1/2\rangle$, i.e. we are at step (b) in Figure \ref{Ben_movie}.
We will now electronically ``tag'' those $\A$ molecules targeted by the global operation $\hat{\A}_f^U$. More precisely, 
we wish to flip the state of the $\A$ inner electronic qubit subject to the function 
$f$, with the control being $\A$'s nearest neighbor $\B$'s inner electronic qubits.
We noted at the end of section \ref{electronic}, 
that many inter-molecular electronic unitaries cannot be achieved with $SU(2)$ pulses. 
However, the case of $f=1$ is simply the operation $\Xi(\B,\A;NOT)$, (see section \ref{global}), 
where one uses 
(\ref{spinthreehalfcnotinner}) for the $C\!NOT$ in the inner electronic qubit subspace. 
The case $f=2$ is more difficult and one must make full use of the construction given in 
(\ref{flips}), to obtain the following pulse sequence 
 \begin{eqnarray}
&&\Xi(\B,\A;\sqrt{NOT})\cdot \{Z^\A_{-\pi/2}\}\cdot \Xi(\B,\A;\sqrt{NOT}) \nonumber\\&&\cdot \{Z^\A_{-\pi/2}\}\cdot\Xi(\B,\A;\sqrt{NOT})\,e^{3i\pi/4}\;\;.\label{flips2}
\end{eqnarray}
where  $\Xi(\B,\A;\sqrt{NOT})\equiv C(S^\B,S^\A;\Sigma )^2$, (see (\ref{mychi})), and
\begin{equation}
C(S^\B,S^\A;\Sigma)\equiv e^{i\frac{1}{2}\pi S_Y^\A}e^{-i\frac{1}{4}\pi (S_Z^\A+S_Z^\B-2S_Z^\A S_Z^\B)}e^{-i\frac{1}{2}\pi S_Y^\A}e^{i\pi/4}\;\;.
\end{equation}
With this pulse sequence one achieves the required spin flip of the $\A$ molecule's inner electronic qubit up to a phase. This global phase
 is unimportant as it will be removed in step (e), when we undo the unitaries. Finally the 
case $f=0$ is dealt with by first flipping all $\B$ inner electronic qubits, then using the $f=2$ operation above, and then 
flipping back. In the case of Figure \ref{Ben_movie}, we have chosen to implement the tag 
$f=1$ operation in step (c).
Once we have set the electronic tag on the appropriate $\A$ molecules we ``undo'' the $SW\!AP$,
 returning $\B$'s quantum information back into its nuclei and  $\B$'s inner electronic qubits 
to the ground state, arriving at step (d) in Figure \ref{Ben_movie}.

\subsection{Implementing $U$}
We now can implement {\bf any} desired unitary on the flagged $\A$ qubits. The  
construction of this operation is well known \cite{Chuangandnielsen}, and consists of 
applying hard pulses \cite{hard}, to {\em all} of the $\A$ and $\B$ and  nuclei interspersed by CNOT 
gates between the inner electronic qubits and nuclear qubits of all the molecules. 
More precisely, the controlled execution of any given unitary operation on the nuclear spin $I$ controlled by the inner electronic qubit $S$ on all the molecules may be written as
\begin{eqnarray}
C(S,I;U)&=& D\, C(I,S;X)\, E\, C(I,S;X)\, F\;\;,\label{unitary1}\\
U&\equiv &e^{i\alpha}R_z(\beta)R_y(\gamma)R_z(\delta)\nonumber\\
&=&e^{i\alpha}DXEXF \;\;.\label{unitary2}
\end{eqnarray}
In (\ref{unitary1}), $\beta,\;\gamma$, and $\delta$ are the Euler angles of the desired nuclear $SU(2)$ operation, $C(I,S;X)$ is the Hyperfine control-NOT given in (\ref{spinhalfthreehalfscnotinner}), the nuclear rotation operators are given by $R_\delta(\theta)\equiv \exp(-i\theta\,\delta /2)$, where $\delta= x, y, z$ and standard pulse sequences for these  spin rotations can be found in \cite{Levit}. Further $D\equiv R_z(\beta)R_y(\gamma/2)$, 
$E=R_y(-\gamma/2)R_z(-(\delta+\beta)/2)$, $F=R_z((\delta-\beta)/2)$, $X\sim NOT$, and $\alpha$ 
is a phase. Crucially one can  show that $DEF=I$.
From this one can see that the action of the sequence (\ref{unitary1}), on the nuclear spin is unity when the inner electronic qubit is not set. When the inner electronic qubit is set however, the action of (\ref{unitary1}) is to execute $U$ from (\ref{unitary2}) on the nuclear spin up to the phase $\alpha$.
 One then finally applies the local unitary $\exp(i\alpha(S_z^\A+S_z^\B))$, to all 
molecules to yield the final phase factor $\alpha$ in (\ref{unitary2}). 
With this one has the very powerful capability of applying {\em any}
 desired single-qubit unitary to the nuclear qubits of the tagged $\A$ molecules. However, the 
typical operation in the QCAQCA is $\hat{\A}_f$, and this can be executed more simply by
 a Hyperfine $C\!NOT$ between the electronic and nuclear qubits of all the molecules. In Figure \ref{Ben_movie}, we illustrate the application of $\hat{\A}_1$ in subfigure (e).

\subsection{Undoing the unitaries}
Following the arguments made in section \ref{NEQCA} above, we execute step (e), that is: once we have implemented the desired unitary on the $\A$ nuclear qubits we ``undo'' 
steps (b), (c) and (d). We $SW\!AP$ the nuclear qubit into the electrons on all $\B$ molecules, 
undo the tagging of the $\A$ molecules, and then $SW\!AP$ the quantum information back into the
 nuclei of the $\B$ molecules. The system is left with all the inner electronic qubits in 
their ground states while the nuclear qubits have received the global operation $\hat{\A}_f^U$. 

\subsection{Gate duration}
As we noted above, the resulting pulse sequence is quite lengthy. However it may be possible
 to compress many of the above operations. Also all nuclear pulses are fast hard pulses of
 little duration \cite{caveat}, while the primary slowdown arises from the nuclear-electronic
 $SW\!AP$ whose duration is limited by the value of the Hyperfine coupling constant of
 \Nc60 ( $\sim 20$MHz), and the separation of the spectral lines of the $\A$ and $\B$
 molecules in the ESR spectrum of Fig. \ref{fig1} ($\sim 50$MHz). The typical QCAQCA 
global operation, $\hat{\A}_f$, entails a pulse sequence in which the terms 
$[+a_zb_z]$, occurs 15 times, each with an average duration of $50ns$, thus 
roughly bringing the cycle time of the global operation down to $1\mu s$. As noted above, 
the simplest quantum logic gate in the QCAQCA scheme requires $\sim 30$ global operations. 
This finally brings the cycle time for logical gates in the resulting QCAQCA to be 
roughly $\sim 30\mu s$. 

We noted in section \ref{endos}, one has $T_1\sim 1s$ at $T=7^\circ K$, and 
$T_2\sim 20\mu$s for concentrated samples of \Nc60 Group-V endohedral 
 material. As mentioned above, 
recent experiments have indicated a dependence on concentration for $T_2$ and it is commonly  expected in all current spin-based quantum computer implementations 
that $T_2$ will rise towards the value of $T_1$ in the limit of zero spin concentration. 
If $T_2\sim 1$s 
then one can expect on average $10,000-30,000$, logical operations within this dephasing 
time (figure of merit). Indeed, at temperatures lower than $7^\circ$K one might be able to
 achieve far greater figures of merit as the decoherence and dephasing times increase. 
A figure of merit of $10^4$ compares very favorably with most alternative proposed
 implementations for quantum information processors. 

\section{Readout}
One of the most challenging aspects of any solid-state implementation is that of readout. 
We first discuss the possibilities for an ensemble readout for the QCAQCA and later a 
single-spin readout for a Type-I implementation. 

In the original scheme of Lloyd 
\cite{Lloyd93}, the only spin sites which could be individually addressed were those at either
  end of the chain as these spins were frequency differentiated by having a single 
neighbor. This is also the situation here. However, if these two sites were the only 
readout sites on the chain then the architecture would not be scalable in the presence of noise.
 One ensemble readout possibility would be to attach an electrically isolated paramagnetic 
adduct to every $\B$ molecule, ( Gd@C82 may be a possibility). Such readout sites would 
typically possess short relaxation times and one would have to isolate such sites
 from the operation of the primary processor using ESR pulses. It should be possible 
within the cellular automata architecture to selectively readout the state of a single qubit, 
that is, to transfer the state of a given logical qubit into the readout site.
This can be done by adapting the single-qubit 
unitary operator $U$, pulse scheme in \cite{Benjamin99}. This pulse sequence is arranged so that
 a logical qubit (which is usually encoded in four spins, $\A\B\A\B$), gets driven into
 the state of a single spin which is then subject to a unitary operation, $U$, through the application of
the global operation (say $\hat{\B}_f^U$, when the qubit has been forced into the spin state of a $\B$ molecule). The spin neighborhood of this target $\B$ spin is arranged so that this global operation only effects the state of that spin alone with an appropriate choice for $f$. 
For readout one could instead apply an operation $\hat{\B}_f^{SW\!AP}$, 
which swaps out the state of the  spin into the readout site. This would be read and
 then $SW\!AP$ed back.
 
For a Type-I quantum information processor based on the $\A\B\A\B$ chain of \Nc60 and
 \Pc60 one must be capable of performing a single electron spin readout. We 
suggest a number of possible technologies that may be capable of performing such a
 readout below. Before mentioning these we return to section \ref{neswap} and Equation 
(\ref{spinhalfthreehalfswap2}). There we noticed that this pulse sequence was able to 
$SW\!AP$ the qubit between the highest and lowest weight subspaces of the nuclear and 
electronic spins, $I=1/2$ and $S=3/2$, and in the process increase the detectable
 magnetic signature by a factor $\sim 1000$. 

We will now further enquire whether there exists
 similar pulse sequences which could further $SW\!AP$ out the quantum information now
 stored in the outer electronic qubit into a separate coupled electronic system of larger spin. This
 could be very advantageous, for example, for the purposes of coupling into the spin-chain,
 a nano-molecular magnet with a spin of $\sim 10-30$ as a potential readout site \cite{sessoli}.
 As the $SW\!AP$ gate is built from $C\!NOT$ gates we can focus on the latter. Further, 
the pulse sequences must  only consist  of higher spin representations of the Pauli group. 
There may be many such pulse sequences and we here present one set of sequences. 
The ones given below perform a $C\!NOT$ between the highest and lowest weight subspaces 
of spins $I$ and $S$. We set $I=a/2$ and $S=b/2$, where $b\ge a$ and both $a$ and $b$ are 
odd integers. Letting $\alpha=[a/2]$ and $\beta=[b/2]$, where $[]$ is the integer part of 
the fraction, we can find $C\!NOT$ pulses for the following four cases:
\begin{widetext}
\begin{eqnarray}
\alpha\;\;{\rm even,}\;\;&\beta\;\;{\rm even:}\;\;& C_(I,S;NOT)=[-S_y][- S_z- I_z][+2S_zI_z][S_y]e^{+ i\pi/4}\;\;,\\
& &C_(S,I;NOT)=[-I_y][- S_z- I_z][+2S_zI_z][I_y]e^{+ i\pi/4}\;\;,\\
\alpha\;\;{\rm odd,}\;\;&\beta\;\;{\rm odd:}\;\;& C_(I,S;NOT)=[-S_y][+ S_z+ I_z][+2S_zI_z][S_y]e^{+ i\pi/4}\;\;,\\
& &C_(S,I;NOT)=[-I_y][+ S_z+ I_z][+2S_zI_z][+I_y]e^{+ i\pi/4}\;\;,\\
\alpha\;\;{\rm odd,}\;\;&\beta\;\;{\rm even:}\;\;& C_(I,S;NOT)=[S_y][+ S_z+ I_z][+2S_zI_z][-S_y]e^{+ i\pi/4}\;\;,\\
& &C_(S,I;NOT)=[I_y][- S_z- I_z][+2S_zI_z][-I_y]e^{+ i\pi/4}\;\;,\\
\alpha\;\;{\rm even,}\;\;&\beta\;\;{\rm odd:}\;\;& C_(I,S;NOT)=[S_y][- S_z- I_z][+2S_zI_z][-S_y]e^{+ i\pi/4}\;\;,\\
& &C_(S,I;NOT)=[I_y][+ S_z+ I_z][+2S_zI_z][-I_y]e^{+ i\pi/4}\;\;.\\
\end{eqnarray}
\end{widetext}

A Type-I quantum computer design will require the capability of single qubit readout 
and this translates in our case to the capability of reading out the spin state of the endohedral electrons.  
One can classify the various  readout methods into the general 
categories (i) force, (ii) electric, and (iii) optical, measurement techniques.  
As was mentioned before, since the electronic wavefunction of the trapped atom is totally
 confined within the C60 cage, spin measurement techniques that involve the physical
 transport of this spin outside the molecule are not possible here. Some techniques, such as 
ODMR (optically detected magnetic resonance), have so far not proved possible with 
 \Nc60 and \Pc60, and thus the addition of separate readout sites which can 
be dynamically coupled and decoupled from the primary processor via NMR/ESR pulses are warranted. 

Possible techniques in the above mentioned categories include (i) Magnetic
 Resonance Force Microscopy (MRFM) \cite{MRFM}; (iia) Micro-Squids, a technology which
 is already capable of discriminating a $\Delta m_S=30$ \cite{Pakes}; (iib) 
Scanning-Tunneling-Microscope ESR, a technique which, though not well understood,
 has yielded single molecule ESR spectroscopy of iron impurities in Silicon 
\cite{STMESR}, and in single BDPA molecules on Silicon \cite{Durkan}. (iic) A
single molecule endohedral fullerene single electron transistor \cite{mcewen}. 
This is a very challenging technology 
which has  yielded a single electron tunneling current through a C60 molecule with 
an electro-mechanical coupling to the quantized motion of the entire molecule. Performing such
 an experiment with \Nc60 may yield endohedral spin information; (iiia) Magnetic
 coupling of the endohedral electronic spin to a solid-state paramagnetic optical
 dipole such as a nanocrystaline N-V  center in  diamond. The optical paramagnetic center
can then be probed 
via optical shelving techniques \cite{NV}; (iiib) ODMR via a magnetic coupling to 
a paramagnetic endohedral adduct which is chemically bonded to the \Nc60 molecule.
For use with ODMR the adduct endohedral should ideally  possesses an optical 
transition in the visible spectrum such as ${\rm Er}_x{\rm Sc}_{3-x}$N@C80 \cite{Dorn} . 

The techniques suggested above
 are, in some cases, already at the single electron detection level (though not capable of detecting the spin orientation), while the out-coupling of the quantum information into a large-spin system
 may allow the techniques of micro-Squids and MRFM as they presently stand to act as a 
readout of a single endohedral electronic spin.

\section{Conclusion}
The philosophy taken in this work with regard to the construction of a quantum information 
processor is to follow as closely as possible the system Hamiltonian that nature provides. The globally addressed approach has very significant advantages over the local gating approach. Global addressing 
avoids the buildup of numerous decoherence pathways associated with the effects of the local gates on the processor. Global addressing also avoids the very significant problem of the scaling up of the external resources required to execute local gating, eg. numerous metallic contacts or multi-frequency MW/RF generators with increasing bandwidths. Although it may be true that for local gating, the external resources required increase polynomially with the number of qubits in the processor, such an overhead may not be experimentally feasible as the size of the processor grows large.

 In some sense the technical complexities involved
 in the construction and operation of local gating are transferred into the  quantum cellular
 automata ``software''. As shown above, the detailed operation of a quantum cellular automata 
quantum computer architecture does not follow the standard quantum circuit model. 
This should not be seen as a disadvantage as all quantum circuit algorithms can be ``compiled'' 
to run on a QCAQCA with a polynomial overhead \cite{Benjamin99}. Indeed the QCAQCA may 
be able to run programs that do not follow the quantum circuit model (see \cite{Raussendorf} 
for an example of a non-circuit quantum program in another architecture). Furthermore, 
as the fundamental quantum operations are implemented via the very well developed methods 
of NMR and ESR, the fidelity of software execution is heightened. The QCAQCA also has the 
very important property that it can execute quantum computations in parallel, a 
characteristic necessary for the ultimate scalability of the architecture. 

In  this work we considered the endohedral fullerene material \Nc60 and \Pc60 and 
argued that the system Hamiltonians and interactions present are sufficient to implement 
universal quantum computation via a quantum cellular automata architecture. Following the 
DiVincenzo criteria: (i) we argued that the Group-V endohedral materials behave as 
essentially electrostatically isolated nanoscopic molecular neutral atom traps that 
can be physically manipulated with relative ease using STM techniques and perhaps may 
be self-assembled using fullerene supra-chemistry. (ii) the electronic decoherence 
times are perhaps the longest seen in any molecular system, of the order of seconds at 
temperatures of $\sim 7^\circ$K, which is also an upper limit for the electronic 
dephasing time. The nuclear decoherence times have not yet been measured but they are 
expected to be several orders of magnitude longer than the electronic decoherence times. 
(iii) The complete polarization of both the nuclear and electronic computational spins is 
feasible as half of the spins (electrons), within the entire system are completely polarised 
at moderate conditions of temperature and magnetic field strength. There are efficient cooling 
schemes exist to shuffle the unpolarized spins away from the computational spins. 
(iv) entanglement can be generated via the inter-molecule magnetic dipole interaction, 
the strength of which has been measured to be $\sim 50$MHz. Armed with the two species
 \Nc60 and \Pc60, we showed that one has sufficient control to implement a 
two-component quantum cellular automata quantum computing architecture storing the quantum information in the nuclear spins while using the electrons as a local bus. We further found that the expected ``figure of merit'' compares very well with other proposed solid-state quantum computer designs. We also discovered pulse sequences that $SW\!AP$ed the quantum information between the highest and lowest weight spaces of two spins of different size, e.g. $S=1/2\leftrightarrow S=11/2$. These sequences could prove useful for out-coupling the quantum information into a spin readout system with large spin.

DiVincenzo's final criterion: (v) an  efficient single qubit readout, is perhaps the most difficult challenge for any solid-state based quantum computer design. We have proposed various possible ensemble and single-issue (Type-I), readout technologies many of which are themselves the subjects of intense study.  The ideas presented here combine the tremendous power of NMR and ESR science together with the very ``clean'', and almost atomic, systems presented by the endohedrals \Nc60 and \Pc60. These, married with supra-molecular chemistry, may provide a very real possibility for a physical implementation of a quantum information processor.

\section*{Acknowledgments}
The author thankfully acknowledges useful discussions with Dieter Suter and Wolfgang Harneit.
 This work was supported by the EU IST FET QIPC project QIPD-DF (http://planck.thphys.may.ie/QIPDDF).

\newpage
\begin{table}[H]
\begin{center}
\begin{tabular}{||l|c|r|r||}
\hline\hline
$B_z=2T$ & &{\bf $A={}^{15}${N@C60}}  & {\bf $B={}^{31}${P@C60}}   \\ \hline
Electronic Zeeman Energy & $g_e\mu_eB_z$ & 56GHz & 56GHz\\\hline
Nuclear Zeeman Energy  & $g_N\mu_NB_z$ &  -6.1MHz & 34.5MHz\\\hline
Hyperfine Coupling Constant & $A$ & 21.2MHz & 138.4MHz\\\hline
\hline\hline
\end{tabular}
\caption{Table of Hyperfine coupling constants and interaction energies for interacting \Nc60, \Pc60 molecules  at $B_z=2T$.}
\label{table1}
\end{center}
\end{table}


\begin{thebibliography}{99}
\bibitem{hard}Pulse sequences are said to be ``hard'' when the RF power is large enough to allow the desired pulses to be of very short duration. Such hard pulses are broadband and excite a range of frequencies \cite{Levit}.
\bibitem{caveat}It may be however, that the attainment of hard nuclear pulses on the scale required above is beyond current hardware capabilities. An alternative is to substitute a multi-frequency soft pulse for the hard nuclear pulse. This will take slightly longer to implement depending on how "soft" the pulse is but this may be improved with pulse shaping.
\end{thebibliography}
\end{document}